\documentstyle[twocolumn,prl,aps]{revtex} 
\begin{document} \parskip =6truept
\pagestyle{empty}

\title{$\hbox{$~$\hskip 5.8truein$~$}\above1pt \hbox{$~$\hskip2truein$~$}$
\hfill {\it May 1997}
\Large \\ \vskip 10truept
Los Alamos National Laboratory 
\hfill $~$\\
Theoretical Division Special Feature 
\hfill $~$\\ \vskip -18truept
\hbox{$~$} \hskip .5truein $\hbox{$~$\hskip2.5truein$~$}\above1pt \hbox{$~$\hskip2truein$~$}$
\hfill $~$\\ \vskip 10truept
T-1 
\hfill $~$\\
Equation of State \& 
\hfill $~$\\
Mechanics of Materials 
\hfill $~$\\ \vskip -18truept
\hbox{$~$} \hskip .5truein $\hbox{$~$\hskip5.3truein$~$}\above1pt \hbox{$~$\hskip2truein$~$}$
\hfill $~$\\ {\bf 
Modeling Metallic Microstructure Using Moving Finite Elements
} \hfill $~$ \normalsize \\
Andrew Kuprat and J. Tinka Gammel
\hfill $~$} 
\maketitle

To attack the general problem of evolution of metallic grain
microstructure in 3D, we employ two different methods: using Monte
Carlo (MC) techniques to anneal an effective discrete model on a fixed
lattice [1], and using Gradient Weighted Moving Finite Elements (GWMFE)
to evolve grain boundaries by mean curvature (local velocity
proportional to the local curvature) -- the simplest continuum model
for grain evolution [2,3].  These two models reduce to each other in
the appropriate limits.  MC has the advantage of computational and
algorithmic simplicity while GWMFE is more easily adaptable to
including coupled transport fields such as vacancies and impurities,
and other geometrical and physical complexities occurring in the
simulation of interconnect-via interactions as related to
electromigration reliability of sub-micron integrated circuits.  Cross
comparisons of the two calculational techniques will produce the best
synthesis of accurate physical models and computational speed.  We also
compare to other methods, such as 2D front-tracking [3].

As an example of MC grain evolution in the confined geometry of an Al
interconnect on a semiconductor chip,\break
\vfill
Fig. 1. 
Mean grain size normalized to the total volume as function of time and
number of neighbors N from Monte Carlo evolution of the Potts model [1]
on an unstructured grid generated by the LANL X3D grid code.  Note
growth stagnates for N$<$5.
\eject
\noindent
we have evolved an initial
microstructure of randomly oriented grains on a fixed lattice using a
discrete effective classical spin model originally developed to model
magnetic domains (Potts model) [1].  
For the first few time steps, MC
scales with the number of volume points. Activity binning and N-fold or
cluster flip MC techniques allow significant speed-ups in the time
evolution [1]. (Incorporation of volume information may reduce the
amount of these speed-ups.) We evolved both a regular fcc lattice,
where discreteness trapping is not a problem [1], and unstructured
grids resulting from the LANL X3D code, where we see from Fig. 1 that
5th neighbor interactions are needed in order to get normal grain
growth without pinning at finite grain size.  We are investigating how
various algorithms for treating further neighbors effect these MC grain
growth statistics.  Having the ability to use the MC code on the X3D
generated geometries and meshes will greatly enhance it's usefulness as
a tool to generate initial conditions for the GWMFE code.

In the finite element approach, we represent the metallic grains on an
unstructured tetrahedral mesh, generated using the LANL X3D grid code.
We use an implicit implementation of the GWMFE method to move the grain
surface (interface) triangles.  Although volumes are deformed by the
moving grid, the computational complexity of the method is only 2D, not
3D, because GWMFE moves triangles, not tetrahedra [2].  Fig. 2a shows
an intermediate microstructure, with the average grain size
approximately half the width, resulting from MC evolution on an
underlying fcc grid, which we used as input to GWMFE.  After a few
GWMFE time steps, the initially jagged interfaces between the grains (a
consequence of the fixed lattice used in the MC simulation) have begun
to smooth out under the action of mean curvature motion (Fig. 2b).  In
Fig. 2c, GWMFE has evolved the microstructure to a completely smooth
state.  Visible in the interior, due to one of the foreground grains
being made ``semitransparent'', are ``triple lines'', where three
grains meet at 120$^\circ$ angles, and ``tetrahedral points'' where
four materials meet.  
On the surface of the geometry, we see that the 
boundary constraint leads to several triple points (analogous to 2D
simulations) where three materials meet at 120$^\circ$ angles.  All
interface angles agree with predictions for mean-curvature motion.
However, the underlying grid (Fig. 2d) composed of interface triangles
is becoming distorted, and must be refined in some areas and derefined
in other areas.  Although the X3D code can accomplish the necessary
topological changes in a semi-automated manner at this time, we are
currently completing development of a fully-automated topological
analyzer called ``Graph Massage'' which will seamlessly accomplish the
necessary topological changes for the simulation to continue while
preserving grain interfaces.  High level routines detect major
topological ``events'' (such as the collapse of an entire grain, which
is then simulated correctly by restricting mesh connectivity changes to
only those that sufficiently preserve material volumes).  We next plan
to investigate grain boundary evolution under isotropic thermal strain
starting from initially randomly oriented grains.

To add the capability of correctly modeling grain boundary evolution in
the environment of time-dependent stress and strain due to unsteady and
spatially nonuniform temperature and impurity gradients, it will be
necessary to couple the GWMFE code to a truly 3D physics code.
Currently under development for this purpose is the Arbitrary
Lagrangian Finite Element (ALFE) code which is a fully-implicit
Galerkin Finite Element method for tracking the evolution of
fully-coupled systems of partial differential equations.  The code
under development will use the latest ``matrix-free'' iterative
methods, and thus will be highly parallelizable.  The fully-implicit
adaptive time-stepping scheme used will allow for efficient resolution
of phenomena at various time-scales, while access to the X3D adaptive
grid refinement capabilities will allow ALFE to track phenomena at
various spatial scales as well.  Finally we note that ALFE is an
``Arbitrary Lagrangian'' code, which means it has the capability of
correctly tracking the evolution of phenomena in an arbitrary,
spatially and temporally nonuniform, coordinate system. This is crucial
for the grain growth simulation, where volume tetrahedral movement is
``slaved'' to the interface triangle movement computed by GWMFE.

We are also currently evaluating a 3D electromigration simulation code
from Motorola which will take as input the microstructure from Fig. 2d
and predict electromigration reliability for the interconnect being
simulated.  These studies will establish whether the additional
overhead of a fully 3D model is justified, or if 2D simulations [3]
adequately describe the evolved microstructure.  We expect that 3D
models will be important in intrinsically nonplanar geometries, such as
vias.
 
\vskip 12truept\noindent
[1] E.A. Holm, Ph.D. Thesis, University of Michigan, 1992, unpublished. 
\hfil\break
[2] Andrew Kuprat, in the Proceedings of the Fifth International
Conference on Numerical Grid Generation in Computational Fluid Dynamics
and Related Fields, Mississippi State University, Mississippi, 1-5
April, 1996.
\hfil\break
[3] H.J. Frost and C.V. Thompson, J. Elec. Mater. 17, 447 (1988).

\vfill
Fig. 2.  
GWMFE evolved microstructure as a function of dimensionless time
(surface tension of order 1 and spatial dimensions of order 1):
(a) t=0: MC model evolved microstructure used as input.
(b) t=0.1: GWMFE has smoothed the initially jagged interfaces.
There is otherwise no significant grain boundary motion.
(c) t=30: The smooth boundaries allow use of large timesteps in the
implicit adaptive GWMFE scheme. 
(d) Underlying surface grid of triangles at the same time as in (c).
(Volume tetrahedra are obscured.)  Topological change is necessary at
this point.

\end{document}